\documentclass[preprint,showpacs,preprintnumbers,amsmath,amssymb,showkeys]{revtex4}

\usepackage{graphicx}
\usepackage{color}
\usepackage{dcolumn}
\usepackage{bm}
\usepackage{amssymb,amsmath}

\begin{document}


\title{Microscopic Irreversibility and the H Theorem}

\author{Jose A. Magpantay}
\email{jose.magpantay@up.edu.ph} \affiliation{National Institute
of Physics and Technology Management Center, University of the Philippines, Quezon City,
Philippines\\}

\date{\today}

\begin{abstract}
Time-reversal had always been assumed to be a symmetry of physics at the fundamental level. In this paper we will explore the violations of time-reversal symmetry at the fundamental level and the consequence on thermodynamic systems. First, we will argue from current physics that the universe dynamics is not time-reversal invariant. Second, we will argue that any thermodynamic system cannot be isolated completely from the universe. We then discuss how these two make the dynamics of thermodynamics sytems very weakly irreversible at the classical and quantum level. Since time-reversal is no longer a symmetry of realistic systems, the problem of how macroscopic irreversibility arises from microscopic reversibility becomes irrelevant because there is no longer microscopic reversibility. At the classical level of a thermodynamic system, we show that the H Theorem of Boltzmann is still valid even without microscopic reversibility. We do this by deriving a modified H Theorem, which still shows entropy monotonically increasing. At the quantum level, we explicitly show the effect of CP violation, small irreversible changes on the internal states of the nuclear and atomic energy levels of thermodynamic systems. Thus, we remove Loschmidt's objection to Boltzmann's ideas.     
\end{abstract}

\maketitle

\section{\label{sec:level1}Introduction}
	How macroscopic irreversibility arises from microscopic reversibility, is a long standing problem of physics, which, although explained by Boltzmann towards the end of the 19th century (for a clear, modern presentation of Boltzmann's ideas, see \cite{Lebowitz}) was still given various explanations during the past thirty years or so. Two of the ideas that attracted attention, within classical physics, are (i) transient fluctuation theory, which posits causality as the cause of the breakdown of reversibility \cite{Evans}, and, (ii) chaos theory, which was argued by some as inherently irreversible because the probability distributions of these processes spread out \cite{Prigogine}. These ideas have their own shortcomings - causality as a cause of irreversibility borders on tautology as pointed out in \cite{Pietsch} while chaotic processes, described in terms of trajectories, strictly speaking are also reversible as argued in \cite{Bricmont}.   
	
	Why is Boltzmann's solution still questioned in spite of the fact that irreversible phenomena like non-equilibrium thermodynamics (Onsager relations), diffusion (Fick's law), and to a certain extent, hydrodynamics (Navier-Stokes equations), can be derived from Boltzmann's theory as discussed already in textbooks \cite{Huang}, \cite{Reichl}? The reason is the exact time-reversal invariance of the underlying particle dynamics. For the critics, statistical arguments (higher entropy states have bigger phase spaces thus more probable) and initial condition (system started with low entropy) cannot rule out macroscopic processes with decreasing entropy because of the particle dynamics' time-reversal symmetry. This is essentially Loschmidt's criticism of Boltzmann's proof. 
	
	But how valid really is time-reversal symmetry in classical physics? Newtonian dynamics, with its Galilean space-time (absolute time) is invariant under t $\rightarrow$ -t for non-dissipative particle dynamics (most have velocity independent forces, for magnetism, the reversal in velocity is accompanied by reversal in the field). An underlying assumption in considering thermodynamic systems is that they can be completely isolated from the environment or the rest of the universe. In Sextion II, we will first present arguments, from current physics,  that the universe's dynamics is not time-reversal invariant.  Then we will argue that any thermodynamic system cannot be isolated completely from its environment or the rest of the universe. In Section III, we will show that this interaction with the universe invalidates time-reversal symmetry for such a system. We will do this by showing that the classical equations of motion is no longer time-reversal invariant, but very weakly, due to the interaction with the environment. Since we no longer have microscopic reversibility, we then investigate what happens to the H Theorem of Boltzmann. This is important because Boltzmann's proof of the H Theorem rests on microscopic reversibility. We then show that even if time-reversal invariance has been broken (rather weakly) at the microscopic level, the system's entropy will still increase, i.e., the H Theorem is still valid. In effect, what we have is Boltzmann's proof of entropy increase without Loschmidt's objection because there is no longer microscopic reversibility. 

	In  Section IV, we discuss time-reversal symmetry at the quantum level. We argue that even quantum mechanically, Loschmidt's objection cannot still be overruled if we assume time-reversal symmetry at the fundamental level. We then propose that phase transitions, inflationary expansion and the breakdown of CP symmetry - processes during the universe's early history - introduced irreversibility in fundamental physics. Of the three, it is CP violation that has calculable effect on violation of time-reversal symmetry in thermodynamics systems. This is what we do in Section IV where we discuss the sources of CP violation from the fundamental forces and the resulting electric dipole moments of the neutron and electron. Then we compute the perturbations on nuclear and atomic energy levels due to the small induced electric dipole moments. Although the effect of these perturbations are very small, the changes they introduce violate time-reversal symmetry. These in turn will overrule Loschmidt's objection because the internal states of the thermodynamic system have built-in irreversibility.
	        	
	In essence, what we will show is that all physical systems are irreversible, they cannot retrace back their trajectories (reversibility implies that as time flows, all the particles reverse velocities, get back to the starting positions and reverse again the velocities to get to the exact initial conditions) but they may recur (get back to the initial positions and velocities without reversing velocities). To elucidate this point, let us answer the question - why do we not feel uneasy watching a video of two colliding billiard balls run backwards if there is no longer time-reversal symmetry? There are two reasons - (1) recurrence, i.e., the reverse sequence of the collission can happen (if we play the game long enough) and may have happened in the past (in some other games), (2) when we play the film backwards, which happens in the forward direction in time, it is only the film sequence 'moving back in time' so to speak and not the entire universe. In other words, there is no contradiction with microscopic irreversibility and thus the reverse sequence of the collission also seem to be a 'natural sequence of events'. 
	
	Argument 1 is an experiential reason for 'naturalness' of sequence of events. But this naturalness has a caveat, which reflects the irreversibility of the microscopic dynamics. This will be discussed again in the conclusion after microscopic irreversibility had been established in the next sections. 
	
	Argument 2 begs the question why the entire universe, which is the only true isolated system, cannot run in reverse sequence. The reason has to do with fundamental physics - current ideas in elementary particles and cosmology will show that the universe's evolution is not time-reversal invariant. And this will also lead to the breakdown of time-reversal invariance of thermodynamic systems.

\section{\label{sec:level2}Arguments for Microscopic Irrevesibility}
	The ideas on cosmology and standard model of elementary particles that are presented below are now mostly accepted and discussed even in textbooks such as \cite{Ryden}. Thus, we will no longer cite the original literature.
	
	Our present understanding of cosmology points to the universe suddenly coming into existence in a Big Bang from a primordial atom around 13.7 B years ago to give a space-time of dimension 3+1 (at least) and a soup of fundamental particles. This clearly hints of irreversibility because even if the fundamental law(s), which we do not know yet, that govern(s) the very early universe (earlier than $10^{-43}$ second) is(are) time-reversal invariant, the evolution was from a simple system (the primordial atom) with very low entropy to a more complicated system (with 3+1 dimensions and soup of quarks, leptons, vector bosons, gravitons and whatever particles that may have existed then) with much higher entropy . 
	
	Since the universe at $t < 10^{-43}$ second did not reverse back to the primordial atom, the subsequent developments during the still very early phases of its life further sealed the universe's irreversible evolution. First, phase transitions (spontaneous symmetry breaking and we presently know there are three but we could not rule out others, for example, if supersymmetry plays a role in fundamental particles, then it must also be spontaneously broken at some time scale), led to the separation of the forces with gravity decoupling first at $t = 10^{-43}$ second, then the strong force decoupling from the electro-weak forces at $t = 10^{-35}$ second and finally the weak decoupling from the electromagnetic at $t=10^{-12}$ second. These phase transitions are time-irreversible because the evolution is from more symmetry to less. There is no known case yet of spontaneous symmetry generation by an isolated system, i.e., a physical system undoing its choice of direction for the ground state. Second, CP violation locked the universe to having more matter than anti-matter thus allowing the creation of protons, neutrons and electrons instead of having a universe of photons. And by the CPT theorem, this means the time-reversal symmetry was explicitly but weakly broken. Third, the inflationary expansion for a very brief moment, from $t = 10^{-35}$ to $10^{-32}$ second of the universe's existence was triggered by the rolling down from a false vacuum of higher energy to a true vacuum of much lower energy, and this resulted in a temperature fluctuations of the order of $10^{-5}$ K when the universe  was around 400,000 years old. This very small temperature fluctuation seeded the formation of large scale structures, the galaxies and clusters of galaxies. Thus, the universe's early dynamics has built into it irreversibility. Finally, with the current ideas on dark energy and accelerated expansion, the universe may end in a Big Freeze or Big Rip. Even if not all of these ideas are exactly correct, it is clear the Big Bang cannot be un-Banged. 

	The breakdown of time-reversal symmetry during the very early phase of the evolution of the universe is reflected in the dynamics of the present macroscopic structures in the universe. And this is clear at various physical scales. Galaxy dynamics - from the formation of galaxies with the early stars to the presence of massive blackholes at centers to collissions of galaxies resulting in mergers that result in bigger galaxies - is clearly time-irreversible. Stellar dynamics - from birth to death of stars, which created elements heavier than helium and can no longer be unmade back to hydrogen and helium - is time-irreversible. Solar system dynamics - from formation of the solar system to eventual death of the sun to a white dwarf and the corresponding fate of the planets - is time-irreversible. Geophysical processes - plate tectonics, volcanic activities, climate changes, weathering and the general 'falling apart' of things - are time-irreversible. And finally biological systems, include experimentalists that measure thermodynamic systems, with its life cycles are definitely time-irreversible processes.
		
	Given the preponderance of time-irreversible processes in the universe, it is rather surprising that we even formed a notion of time-reversal symmetry. Today, the established view is that the fundamental law(s) is/are time-reversal invariant and the myriad of complicated irreversible phenomena that we observe is due to the initial condition (low entropy, not in equilibrium) of the universe. This is rather unusual because physicists tend to give more importance to dynamics than special circumstance (initial condition). We argue that if it is the initial condition of the Big Bang and not dynamics as the cause of all the irreversible phenomena, then the three phase transitions, CP violation and inflation must be accounted for by the initial condition (low entropy and not in equilibrium). This is not clear at all because the very general initial condition could have led to other evolutions of the physical laws - CPT or PT violation instead of CP, forces having different relative strengths instead of the present values or inflation turning off earlier or later than $10^{-32}$ sec. Since the early universe evolved in the way we know it now and not following the alternatives cited here, then dynamics must play an important role.   
	
	The question now is why do we believe in the time-reversal symmetry of the fundamental law(s) (right after the Big Bang up to $10^{-43}$ sec) even though we do not know yet what it/these is/are? There are two possible reasons. The first is that except for the weak interaction which violates time-reversal weakly and the possible violation by the strong force because of the non-trivial topology of QCD, the other two forces we know now appear to be invariant under time-reversal and we extrapolate this to be valid even for the early fundamental law(s). But there is built-in irreversibility in the four fundamental forces because their strengths were determined by the irreversible phase transitions that happened during the universe's early history. Thus, these forces are not really time-reversal invariant. If the four forces now are not strictly time-reversal invariant, we cannot really say that they are invariant when we extrapolate them to right after the Big Bang. The second is that terrestrial phenomena and elementary particles only require Galilean and Minkowskian space-time. Knowing what we know now, that the universe is expanding and even accelerating in its expansion, space-time cannot be Galilean or Minkowskian. Whatever it is, the physical theory that we will formulate in such a space-time cannot be time-reversal invariant for it must flow with the accelerated expansion of the universe. 
	
	In short, the irreversibility of the universe is not farfetched from having a dynamical component and not merely be due to initial condition. We will not make use of these ideas in Section III of the paper where we just need to make use of the universe's irreversibility, regardless of the reason, to show that at the level of Newtonian dynamics, this will lead to microscopic irreversibility.   
	
	The macro-micro problem in thermodynamic systems arises because of the time-reversal invariance of Newtonian mechanics. The second-order differential equation nature of Newton's laws is a given thus time-reversal invariant. If we accept the arguments in the previous paragraphs, then the forces that particles experience now are not really time-reversal invariant. Furthermore, the building blocks of any thermodynamic system, elements like carbon, oxygen, nitrogen, etc., are products of irreversible processes that happened in some stars and as such have built-in irreversibility also. Thus, microscopic dynamics at the level of Newtonian physics strictly speaking is also not time-reversal invariant. 
	
	But even if we neglect these arguments for Newtonian irreversibility, i.e.,  if we assume (1) that the universe's irreversible evolution is due to the initial condition (and the forces are time-reversal invariant), and (2) the elements of thermodynamic systems are Markovian and thus would have lost the memory of the irreversible processes that led to their existence, we argue that the fact that a thermodynamic system cannot be completely isolated from the rest of the universe will force it to march along with the rest of the Universe. Its dynamics will not be time-reversal invariant but very weakly. This is what we focus on in the next section. 
	
	There are a number of ways any thermodynamic system (in our laboratories or elsewhere) will couple, even if very weakly, to the rest of the Universe. First, the universe is bathed with electromagnetic radiation of different frequencies (from gamma rays produced in the nucleus and violent processes near black holes to the microwave background radiation leftover from the Big Bang). Although most of these radiation will not have enough energy to affect thermodynamic systems, the extremely energetic ones, X rays and gamma rays, even with small probability of interacting with protons and electrons will interact with the system because of the sheer number of particles. 
	
	Second, aside from the electromagnetic radiation, the earth is bathed with energetic charged particles (mostly protons, electrons and pions), very weakly interacting neutrinos and maybe some other particles that we have not established yet (WIMPs and other particles in susy and extra dimension theories). Most of these particles will just go through thermodynamic systems (neutrinos for example) but some, like the charged ones cannot be ruled out from having an effect, may be small, on the system. A rather interesting recent analysis showed that up to 1 WIMP can interact with a nuclei of a human body per minute under certain conditions \cite{Freese}. In other words, a thermodynamic system cannot be ruled out from being affected, even very weakly, by elementary particles. 
	
	Third, the gravitational interaction with the rest of the universe, again very small, cannot be shut off from interacting with thermodyanmic systems. Primarily, the gravitational influence is exerted by the earth, moon and sun. A geophysics analysis showed that the earth's gravitational force, as measured by g, on any object comes from 3 sources - the gravitational attraction of the masses, the tidal forces coming from the moon and the sun and the centrifugal force due to the earth's rotation \cite{Volgyesi}. All these change slowly with time - the earth accretes mass from the constant bombardment of rocks from space, the earth's rotation is slowing down, the moon is receding away from the earth very slowly and the sun will eventually become a white dwarf, passing first through a red giant stage. Though these changes to g are small and rather very slow, it shows that in principle, a thermodynamic system is affected by the irreversible geophysical and solar system processes. There is even an argument that part of the change in g may be due to the slow change in time of Newton's gravitational constant \cite{Volgyesi}, an idea first proposed by Dirac \cite{Dirac}, which if true must be due to unkown fundamental physics. 
	
	And finally, the experimenter (us), in preparing and doing measurements on the system, clearly points further to the fallacy of an isolated system. Thus, any thermodynamic system is never decoupled or isolated and thus cannot march separately in time from the rest of the Universe, which marches only forward in time. In other words, the irreversible evolution of a thermodynamic system is due to its interaction with the environment.
	
	But this cannot be the complete story of irreversibility for a thermodynamic system. The phase transitions, inflationary expansion and CP violation, which must have played  major role in the irreversible dynamics of the universe, must also be playing a role in the irreversible evolution of the thermodynamic system. This is what we will explore in Section IV. Here we will look at the changes in internal structure of the components of the thermodynamic system. This means looking into the atom and nucleus and necessarily the analysis will be quantum mechanical.   
	
\section{\label{sec:level3}A Simple Way to Implement Irreversibility}
	In this section, we will consider a simple way to incorporate irreversibility in thermodynamic systems. We will do it in such a way that we can start from mechanical analysis and follow the usual steps to statistical analysis. Also, we will do things non-relativistically to be as close to Boltzmann's ideas as possible. 
	 
	Consider a closed system and environment. Let the particles of the system have coordinates labeled by $q_i$, masses  $m_i$, with i from 1 to n while those of the environment labeled by $Q_A$ and masses $M_A$, with A from 1 to N. Typically, $n\approx 10^{23}$, i.e., Avogadro's number, while $N\approx 10^{88}$, the estimate of the total number of particles in the universe \cite{Smirnov}. Only a tiny fraction of the environment degrees of freedom will interact with the thermodynamic system but we will just put in everything, as a matter of principle. The environment dynamics will only be required to break time-reversal symmetry because of the arguments presented in the previous section. This can easily be implemented by having an odd function in $\dot{Q}_A$ in $V_{env}$ (see below). 
	
	Let us write down the system-environment lagrangian
\begin{equation}\label{1}
L=L_{sys}+L_{env}-\lambda V_{int},
\end{equation}
where
\begin{subequations}\label{2}
\begin{gather}
L_{sys}=\frac{1}{2}\sum_{i=1}^nm_i\dot{q_i}^2 - V_{sys}(q),\label{first}\\
V_{int}=V_{int}(q_i,Q_A),\label{second}\\
L_{env}= \frac{1}{2}\sum_{A=1}^NM_A\dot{Q_A}^2 - V_{env}(Q_A,\dot{Q_A}).
\end{gather}
\end{subequations}
For simplicity, the system-environment interaction, which provides only energy exchange between the system and environment (thus the system is closed but not isolated) is only coordinates dependent, thus invariant under time-reversal just like the system Lagrangian (quadratic in velocities). Thus, the breakdown of time-reversal symmetry is only encoded in the environment dynamics. 
	
	If $\lambda$ is zero, the system decouples completely from the environment and its dynamics has time-reversal symmetry, i.e., the system can reverse back even as the entire universe irreversibly unfolds forward in time. But when $\lambda$ is no longer zero, even if very, very small, the equations of motion for the system and environment are coupled as shown by
\begin{subequations}\label{3}
\begin{gather}
m_i\ddot{q_i}+\dfrac{\partial V_{sys}}{\partial q_i}+\lambda \dfrac{\partial V_{int}}{\partial q_i}=0,\label{first}\\
M_A\ddot{Q_A}-\frac{d}{dt}\left(\dfrac{\partial V_{env}}{\partial \dot{Q_A}}\right)+\dfrac{\partial V_{env}}{\partial Q_A}+\lambda \dfrac{\partial V_{int}}{\partial Q_A}=0.
\end{gather}
\end{subequations}
These are coupled equations that cannot be solved without specific dynamics and impossible to solve even with specific dynamics because of the sheer number of degrees of freedom. But even without specifying the system dynamics and its interaction with the environment, we can still make definite conclusions about the macro-micro problem. 
	
	Since the environment is much, much bigger than the system, we can safely say that we can neglect the effect of the system on the environment dynamics, i.e., we can neglect the last term of equation (3b). Let the solution of the pure environment dynamics be $\tilde{Q}(t)$. This solution will not be time-reversal invariant to reflect the irreversible processes at various macro physical scales discussed in the previous section. The effect of the environment on the system is through $\tilde{Q}(t)$ as substituted in the last term of equation (3a). Thus, the system dynamics explicitly, although very weakly violates time-reversal symmetry through $\tilde{Q}(t)$. Time-reversal symmetry is now violated even at the microscopic level.

	Before we continue with our analysis, let us be reminded of how the environment is usually treated. The environment is thought to exert random forces on the system and drive the system to equilibrium \cite{Jaynes}. This view is not taken here because although the environment is much, much bigger, how it interacts with the system is still based on fundamental (causal) forces. And since the environment degrees of freedom exhibit irreversibility, its causal forces on the system will also lead to a small irreversibility in a thermodynamic system.
	  
	We begin then from the following system Lagrangian
\begin{equation}\label{4}
L=\frac{1}{2}\sum_{i=1}^nm_i\dot{q_i}^2 - V_{sys}(q)-\lambda V_{int}(\tilde{Q}(t),q).
\end{equation}
The violation of time-reversal symmetry can be made very weak through the strength of the system-environment coupling $\lambda$. The irreversibility can also be made very slow through the behavior of $\tilde{Q}(t)$. These two behaviors reflect the discussions in the previous section and also the fact that the thermodynamic system's energy is almost conserved.

	The realization of the environment's influence on the thermodynamic system in terms of a potential needs explanation. Undoubtedly, the interaction of photons and elementary particles that bathe the thermodynamic system with the protons, neutrons and electrons of the system must be relativistic and quantum mechanical. To represent this in terms of a $V_{int}$, which is the interaction of each molecule with coordinate $\vec{q}$ and the environment particles with coordinate $\vec{Q}$, we just have to remember the Coulomb (for photons) and Yukawa (for massive particles) potentials. For example, an energetic photon interacts with an electron of one of the molecules in the thermodynamic system. This photon must have been emitted by another charged particle from somewhere in the universe and it will cause a Coulomb potential (strictly speaking it should be retarded but we are considering the non-relativistic limit) between the emitter from the environment and the electron of the thermodynamic system. As for the strength of the interaction $\lambda$, this must be very weak because as discussed in Section II, very few of the system particles will interact with the environment particles. As an order of magnitude estimate, if for any given time only a few protons, neutron or electrons will interact with the environment particles, we can say that $\lambda \approx 10^{-23}$ to reflect this very rare interaction at the potential level.
	
	In the case of the gravitation interaction, which cannot be shut-off although extremely weak, as viewed from the thermodynamic system, the positions $\vec{Q}$ of distant particles in other macroscopic structures (outside of the solar system) change very slowly in time. The effect of the earth, the moon and the sun is through gravitational acceleration g, which changes very slowly with time. These influences are easily implementable in terms of a potential $\lambda V_{int}$ and these are really small, although the most dominant as the analysis below will show, still resulting in $\dfrac{dH}{dt}<0$.        

	The H function that we use is the negative of entropy (modulo $k_{B}$ and not the entropy density) and is given by
\begin{equation}\label{5}
H(t)=\int d\vec{q}d\vec{p} f(\vec{q},\vec{p},t)\ln f(\vec{q},\vec{p},t),
\end{equation}
with the one-particle distribution function f(q,p) solved from the first of the BBGKY hierarchy
\begin{equation}\label{6}
\dfrac{\partial f_1(q,p,t)}{\partial t}+\frac{p}{m}\dfrac{\partial f_1}{\partial q}+F^{ext}(\tilde{Q}(t),q)\dfrac{\partial f_1}{\partial p}=\int dq'dp'\dfrac{\partial V_{sys}(\left|q-q'\right|)}{\partial q}\dfrac{\partial f_2(q,p,q',p',t)}{\partial p},
\end{equation}
where $f_2$ is the two-particle distribution, which is supposed to be solved in the next level of the BBGKY hierarchy. The second and third terms of this equation gives the drift term of the change in the one particle distribution. The last term will be approximated by Boltzmann in terms of two particle collisions through his Stozzahl Ansatze, i.e., dilute gas with the two particle distribution given by the product of one particle distributions, which in essence truncates the hierarchy. This means replacing the $f_2$ term  by the collision term given by
\begin{equation}\label{7}
\left(\dfrac{\partial f(\vec{q},\vec{p},t)}{\partial t}\right)_{coll}= -\int d\vec{p}'d\vec{p_1}d\vec{p_1}' T(\vec{p}+\vec{p_1}\rightarrow \vec{p}'+\vec{p_1}')\left[f(\vec{q},\vec{p},t)f(\vec{q},\vec{p_1},t)-f(\vec{q},\vec{p}',t)f(\vec{q},\vec{p_1}',t)\right],
\end{equation}
where we put in the vectors to emphasize the 3 D nature of the quantities and dropped the subscript 1 in the one-particle distribution. The transition rate $T(\vec{p}+\vec{p_1}\rightarrow \vec{p}'+\vec{p_1}')$, which is generally determined using quantum mechanics, gives the probability per time for the two incoming particles of momentum $(\vec{p},\vec{p_1})$ to change to momentum $(\vec{p}',\vec{p_1}')$ as outgoing particles, . In Boltzmann's analysis, this transition rate is symmetric because of time-reversal symmetry, i.e., $T(\vec{p}+\vec{p_1}\rightarrow \vec{p}'+\vec{p_1}')=T(\vec{p}'+\vec{p_1}'\rightarrow \vec{p}+\vec{p_1})$. Also, since there is no external force considered, momentum is conserved, i.e., $\vec{p}+\vec{p_1}=\vec{p}'+\vec{p_1}'$.

	In our case, there is a very weak external force coming from the environment, which violates momentum conservation very weakly also, i.e., 
\begin{equation}\label{8}
\vec{p}+\vec{p_1}=\vec{p}'+\vec{p_1}'+2\lambda \delta t \nabla_{\vec{q}}V_{int}(\tilde{Q}(t),q),
\end{equation}
where $\delta t$ is the collision time. Note, the violation is time-dependent as encoded in $\tilde{Q}(t)$. Because of this, the transition rate for the process becomes $T(\vec{p}+\vec{p_1}\rightarrow \vec{p}'+\vec{p_1}'+2\lambda \delta t \nabla_{q}V_{int}(\tilde{Q}(t),q))$. This can be expressed in terms of the transition rate with time-reversal symmetry $\bar{T}$ that would have appeared in Boltzmann's analysis, i.e., 
\begin{equation}\label{9}
T(\vec{p}+\vec{p_1}\rightarrow \vec{p}'+\vec{p_1}'+2\lambda \delta t \nabla_{q}V_{int}(\tilde{Q}(t),q))=\bar{T}(\vec{p}+\vec{p_1}\rightarrow \vec{p}'+\vec{p_1}')+2\lambda \delta t\nabla_{\vec{p}}\bar{T}\cdot \nabla_{\vec{q}}V_{int}(\tilde{Q}(t),q).
\end{equation}
This expression clearly shows the violation in the time-reversal symmetry of the transition rate is very small as determined by two really small parameters - the collision time $\delta t$, which is typically of the order of a tenth or hundredth of a nanosecond \cite{Huang}, and the strength of the system's interaction with the environment $\lambda$, which must be extremely small considering the previous discussions. Thus, we expect that if there is a correction to the H Theorem of Boltzmann, it would be really small.

	Taking the derivative of H, we find that just like in Boltzmann's case, the drift terms do not contribute by making use of divergence theorems and assuming the one-particle distribution function vanishes at the boundaries of space and momenta. The collision term must now make use of equation (8). Following exactly the same procedure as in Boltzmann's proof, we find
\begin{equation}\label{10}
\begin{split}
\dfrac{dH}{dt}&=-\frac{1}{4}\int d\vec{q}d\vec{p}d\vec{p_1}d\vec{p}'d\vec{p_1}'\left[\bar{T}(\vec{p}+\vec{p_1}\rightarrow \vec{p}'+\vec{p_1}')+2\lambda \delta t\nabla_{\vec{p}}\bar{T}\cdot \nabla_{\vec{q}}V_{int}(\tilde{Q}(t),q)\right]\\
&\quad\left[f(\vec{q},\vec{p},t)f(\vec{q},\vec{p_1},t)-f(\vec{q},\vec{p}',t)f(\vec{q},\vec{p_1}',t)\right]\left[\ln\left(\dfrac{f(\vec{q},\vec{p},t)f(\vec{q},\vec{p_1},t)}{f(\vec{q},\vec{p}',t)f(\vec{q},\vec{p_1}',t)} \right)\right]
\end{split}
\end{equation}  

	The product of the two terms at the end is positive. In Boltzmann's case, we will only have the transition rate, which is also positive definite, thus ensuring that the time rate of change of H is negative (thus the entropy is always increasing). In our case, the violation of the time-reversal symmetry is very small and thus the first term is the time-symmetric transition rate that appears in Boltzmann's proof plus a function that is not positive definite but very small as controlled by two highly infinitesimal quantities (the collision time and the system-environment coupling $\lambda$). Thus, we can say that
\begin{equation}\label{11}
\bar{T}(\vec{p}+\vec{p_1}\rightarrow \vec{p}'+\vec{p_1}')>>2\lambda \delta t\nabla_{\vec{p}}\bar{T}\cdot \nabla_{\vec{q}}V_{int}(\tilde{Q}(t),q)
\end{equation}

	Let us give order of magnitude estimates of the correction to Boltzmann's proof. In the case of electromagnetic interaction, the Coulomb attraction and repulsion between a charge in the system and the electric charges in the universe will cancel out. What contributes is the more direct interaction cited previously, i.e., when a photon hits a charge particle in the system. This will provide a force given by $\nabla_{\vec{q}}V$ given by the Coulomb force (we are neglecting retardation effects). Assuming the charge particle in the universe that produced this photon has a distance of say ($10^{11}$, $10^{25}$) m (for example, coming from the sun or a distant gamma ray burst), putting in the numbers, we find $2\lambda \delta t\nabla_{\vec{p}}\bar{T}\cdot \nabla_{\vec{q}}V_{int}(\tilde{Q}(t),q)\approx (10^{-83}, 10^{-111})\left|\nabla_{\vec{p}}\bar{T}\right|$. The correction is indeed very small compared to $\bar{T}(\vec{p}+\vec{p_1}\rightarrow \vec{p}'+\vec{p_1}')$. 
	
	If it is a massive particle that interacted with one of the particles in the system, by Yukawa interaction this will result in a much smaller correction. 
	
	The biggest correction will come from the gravitational interaction. Let us assume the cosmological principle, i.e., the universe is isotropic and homogeneous. The gravitational pull of objects other than the earth, moon and sun on any molecule in the system will cancel out. As discussed in Section II, the relevant force is given by g, which is slowly changing in time. For a typical molecule with mass $10^{-25}$ kg and in this case taking $\lambda=1$ because of the certainty of g acting on the thermodynamic system, the correction then becomes $10^{-36}\left|\nabla_{\vec{p}}\bar{T}\right|$, still definitely a small correction.    

	All this guarantees that $\dfrac{dH}{dt}$ is negative resulting in no change in the H Theorem even without microscopic reversibility, thus negating Loschmidt's objection. In short, we have the best of both worlds - increasing entropy of Boltzmann and time-irreversibility even at the level of Newtonian microscopic dynamics.

	Finally, we comment on the equilibrium distribution. Strictly speaking, the thermodynamic system will never be in equilibrium as long as the universe irreversibly unfolds in time because the system will be subject to a small, even extremely weak external force. This means that the system will continue to increase its entropy, keeping pace with the universe's increase in entropy. However, to the extent that we can neglect the external force of the environment, the system's entropy will attain maximum value with an equilibrium distribution given by the Maxwell-Boltzmann distribution.  		
\section{\label{sec:level4}CP Violation and Thermodynamics Irreversibility}
	In Section II, we cited the irreversibility of the universe at various macro scales - from galaxy dynamics to biological systems - and we argued that since a thermodynamic system cannot be isolated from the rest of the universe, the Newtonian dynamics of the molecules in the thermodynamic system will also be irreversible, albeit very weakly. In Section III, we then calculated the effect of this very weak microscopic irreversibility on the H Theorem and showed that it does not invalidate Boltzmann's proof of entropy increase and in effect remove Loschmidt's objection. In this section, we will now make use of CP violation, which breaks time-reversal invariance explicitly, and determine its effect on irreversibility of a thermodynamic system. We will not discuss the other two fundamental irreversible processes that happened during the universe's early phase, inflationary expansion and phase transitions, because their effect on a thermodynamics system's irreversibility is too far removed. The scales of these phenomena are much too small to affect thermodynamic systems. But to understand if Loschmidt's objection is relevant in quantum theory, we will first discuss how irreversibility shows up in quantum theory even though it is assumed that the fundamental forces are time-reversal invariant.  
	
	There are several mechanisms for irreversibility in quantum theory. The first mechanism is measurement \cite{von Neumann} because it results in the the collapse of the wave function, which is obviously an irreversible process. But this is unsatisfactory because irreversibility essentially is triggered by a classical observer and not by an intrinsic attribute of a quantum system. The wave function collapse can be made a purely quantum phenomena without a need for a classical observer within the so called many Hilbert space (MHS) approach \cite{Machida} but this requires the introduction of an order parameter $\epsilon$ that measures the degree of decoherence with values in the range $(0,1)$ with total collapse only for $\epsilon=1$ and partial collapse for in between values. This will have important consequence on the reversibility of the wave function collapse. 
	
	The second mechanism is a more natural mechanism than the first for it makes use of quantum fluctuations, thus purely internal to the system. The argument is quantum fluctuations result in loss of information, thus increasing entropy \cite{Perez-Mercader}. It is also argued that quantum fluctuations drive the system to a stable classical configuration, which is the state of maximum entropy. 
	
	The third mechanism and the most recent is dynamical \cite{Snoke} and makes use of the deteministic evolution of wave mechanics via time-dependent perturbation theory plus Stozahl Ansatze (collission process with two particles in and two particles out) to derive a quantum Boltzmann equation that exhibits an irreversible evolution of the number of particles to its equilibrium distribution value. Since this proof of irreversibility of quantum systems is based on dynamics, it shows that the evolution to higher entropy is not a statistical effect. 
	
	All the above mechanisms for irreversible flow assume time-reversal invariance at the fundamental level. Quantum mechanically, is Loschmidt's objection still valid, i.e., can a quantum mechanical system flow to states of lower entropy? Of the three quantum mechanisms for irreversibility, it seems that the least attractive one - measurement by a classical observer that leads to collapse of the wave function - is the one that is most stable against reversible flow. But if we make the collapse as a purely quantum effect through the MHS approach, collapse has a continuous range of parameter and those in the intermediate range is not a total collapse, thus may be reversible. 

	As for quantum fluctuations, there is no a priori reason why  quantum fluctuations cannot drive a system out of equilibrium and evolve towards lower entropy. By the very nature of random phenomena, the state of a system is a matter of probabilities, thus lower entropy states only have smaller probabilities but not zero. Even the third mechanism, since it is based on quantum dynamics that makes the average number of particles flow to equilibrium distribution, did not rule out the system moving away from equilibrium - the probability of such an event can only be guaranteed to be very small. In this sense, this is consistent with the second mechanism based on quantum fluctuations.            

	All these show that even in quantum theory Loschmidt's objection cannot be completely ruled out. And maybe this problem can also be solved by irreversibility at the fundamental level. As argued in Section II, the inflationary expansion, phase transitions and CP violation are irreversible phenomena that happened during the very early phase of the universe. Of the three, the one that we can calculate the effect of fundamental irreversibility on the irreversibility of thermodynamic systems is CP violation. The reason for this is CP violation results in an intrinsic electric dipole moment for particles and we know the electric dipole moment contribution to the Hamiltonian of any system. Thus we can quantum mechanically compute the fundamental irreversibility contribution on a thermodynamic system.
	
	But before we discuss the contribution of CP violation to a thermodynamic system's irreversibility, let us discuss first how CP non-invariance appears at the fundamental level. There are three sources of CP violations at the level of fundamental forces. First, from topological considerations, the QCD lagrangian may include the $\theta$ vacuum term, see for example \cite{Coleman}, given by
\begin{equation}\label{12}
\textit{L}_{\theta}=\theta\dfrac{g^2}{32\pi^2}\int d^{4}x F\tilde{F}.
\end{equation}
This extra term is gauge-invariant and provides a possible solution to the $U_{A}(1)$ problem by providing the gluonic contribution to the $\eta$ meson mass. But the extra term breaks CP and leads to an electric dipole moment (edm) for the neutron. The limit to the neutron's edm is $< 2.9 \times 10^{-26}$ e cm \cite{Baker}and this in turn sets a limit to $\theta \approx 10^{-10}$. We can turn the argument around, i.e., assuming the strong force violates CP very weakly as given by the small value of $\theta$, we get the neutron's electric dipole moment as given. We can now make an estimate how the small breakdown of time-reversal by the fundamental strong force will affect a thermodynamic system by computing how the neutron's edm affect the nucleus' structure.

	The small edm of the neutron will cause three perturbations to the energy levels of the nucleons. First, the electric field of the protons will act on the edm and result in a Stark effect perturbation. Second, the dipole-dipole interaction of the neutrons will also shift the energy levels. Third, the edm of the neutron will act on the protons. The nucleon energy levels are given by the shell model using a Woods-Saxon potential and spin-orbit interaction typically giving energy levels at Mev or ten Mev \cite{Meyerhof}. To estimate the Stark effect, we have to make use of a specific charge distribution, the one that has a good fit to the experimental data of the differential scattering cross-section of high energy electrons, the Fermi charge distribution \cite{Hofstadter}. This distribution is given by
\begin{equation}\label{13}
\rho=\dfrac{\rho_0}{1+\exp{\frac{(r-R)}{\delta}}},
\end{equation}
where $R = 1.18 A^{\frac{1}{3}}$ fm is the nuclear radius and $\delta \approx .4 - .5$ fm for $A > 40$ is the surface depth. An elementary physics calculation will relate the constant $\rho_0$ to the total charge of the nucleus Ze, R, and $\delta$; and give the electric field due to this spherical charge distribution at any distance $r < R$ from the nucleus center as
\begin{equation}\label{14} 
\vec{E}=-\dfrac{\rho_0\delta^3}{\epsilon_0}(\frac{1}{r^2})\left\{2\sum_{n=1}^{\infty}(-1)^n\frac{1}{n^3}\exp(\frac{nr}{\delta})+\sum_{n=1}^{\infty}(-1)^{n+1}\left[\frac{2}{n^3}+\frac{2}{n^2}\frac{r}{\delta}+\frac{1}{n}\frac{r^2}{\delta^2}\right]\exp\left[n(\frac{R-r}{\delta})\right]\right\}\hat{r},
\end{equation}
where $\epsilon_0$ is the electric permittivity of the vacuum. This electric field will give the shift in energy levels of a neutron via the Stark effect (Se) as given by
\begin{equation}\label{15}
H_{Se}= -\vec{p}\cdot\vec{E}
\end{equation}
Putting in the numbers, we find that $H_{Se}\approx10^{-11}E_{sm}$, where $E_{sm}$ are the Mev ranged shell model nucleon energy levels.

	Another perturbation to the neutron's energy levels in the nucleus is the dipole-dipole interaction. This is given by
\begin{equation}\label{16}
H_{dd}=\frac{1}{4\pi\epsilon_0}(\frac{1}{r^5})\left[r^2\vec{p_1}\cdot\vec{p_2}-3\vec{p_1}\cdot\vec{r}\vec{p_2}\cdot\vec{r}\right]
\end{equation}
Putting in the numbers, we find $H_{dd}\approx10^{-26}E_{sm}$, much, much smaller than the Stark effect contribution. This perturbation is time-reversal invariant because it is a product of two edms.

	The third perturbation is the effect of the neutron's dipole moment on a proton and is given by
\begin{equation}\label{17}
H_{dp}=\frac{e}{4\pi\epsilon_0}\dfrac{\vec{p}\cdot\vec{r}}{r^3}
\end{equation}
Putting in the numbers, we find $H_{dp}\approx10^{-12}E_{sm}$, which is of the same order as the Stark effect correction. Note, this perturbation also breaks time-reversal symmetry.

	Taken together, the three perturbations will only produce a very small time-reversal symmetry breaking perturbations on the protons and neutrons of the nucleus of the atoms in the thermodynamic system. Thus, although the changes are not noticeable, we can say that the nucleus of the gas molecules are undergoing very small irreversible changes that will not affect Boltzmann's H Theorem just like in the classical case where the irreversible changes are environment triggered.

	Now, let us consider the second contribution to CP violation, the electro-weak part of the Standard Model. The Standard model, with three families, breaks CP symmetry via the Cabibbo-Kobayashi-Maskawa mixing because of one non-vanishing phase, see for example \cite{Collins}. At the level of particle physics, the breakdown of time-reversal symmetry had been established experimentally in kaon decays and the more recent B meson decays. The macroscopic consequence of this small microscopic irreversibility was derived in the thermodynamics of kaons in thermal bath with pions \cite{Aharony}.   
	
	Since we are interested in the effect of fundamental microscopic irreversibility on thermodynamic systems, we need to know the quark's electric dipole moment attributed to the weak CP violation. Unfortunately, because this happens at the three loop level in the Standard Model \cite{Czarnecki}, this dipole moment is very small, about $10^{-33}$ e cm and this in turn gives a neutron edm of about $10^{-32}$ e cm. Assuming this result, the contribution to the three perturbations discussed above in the nuclear shell model calculations will be six orders of magnitude smaller than the above calculations.  
	
	The third fundamental physics contribution to the breakdown of time-reversal symmetry comes from supersymmetry. Here we will see a marked increase in the neutron's edm. The reason for this is the fact that there are many more non-vanishing phases in supersymmetric standard models. The mixing angles, phases and masses of the sparticles have a range of values \cite{Inui} and this gives a range of values for the neutron edm, with an upper value of $10^{-24}$ e cm. If this is correct, the three corrections to the nuclear shell model energy levels will be two orders of magnitude higher than the ones computed above. Unfortunately, this high value for the neutron's edm apparently should have been detectable with the present experimental accuracy. Since the present experimental limit is much lower, there must be more fine tuning of the supersymmetric mixing angles, masses and phases if supersymmetry is able to explain the neutron's edm.
	
	All the above very small contribution to the breakdown of time-reversal symmetry from fundamental physics will result in changes to the nuclei of the atoms in microscopic systems. The small perturbations will change the nuclear state, although, almost imperceptibly, they definitely will rule out Loschmidt's objection because we have lost time-reversal symmetry.    
	
	The last contribution to the breakdown of time-reversal symmetry in thermodynamic systems from fundamental physics is at the atomic level. The experimental limit on the electron's edm is $< 10^{-28}$ e cm \cite{Hudson}, two orders of magnitude smaller than the neutron's. This will result in a Stark effect correction to the atomic energy levels $E_{a}$. According to Bohr and Schrodinger, $E_{a}$ is at the ev level. The Stark effect contribution is of the order of $10^{-19}E_{a}$. This is much, much smaller than the time-reversal violating corrections in the nucleus. This correction will even be much smaller, about $10^{-30}E_{a}$ if the Standard Model is correct for it predicts an electron edm of around $10^{-39}$ e cm \cite{Pospelov}. 
\section{\label{sec:level5}Conclusion}
	In this paper, we argued the case for microscopic irreversibility for all physical systems at two levels. At the level of Newtonian dynamics, we made use of the fact that a thermodynamic system cannot be isolated from the rest of the universe and the universe's classical dynamics is irreversible to show that the thermodynamic system will go with the irreversible flow of the universe. This led to a modification of the H Theorem as shown in equation (14), which still shows entropy increasing but without microscopic reversibility, thus negating Loschmidt's objection to Boltzmann's ideas.
	
	Second, at the level of quantum mechanics of the nuclei and atoms that make up the thermodynamic system, we consider the effect on the energy levels of the nucleons and the atom from CP violation that comes from three fundamental physics - strong interaction, weak interaction and supersymmetry. We find that the time-reversal violating perturbations will minutely change the internal state of the elements of the thermodynamic system irreversibly.     
	
	Now we comment on the caveat hinted in the introduction regarding the collision of two billiard balls. If there is microscopic reversibility, the reverse sequence of the collision is really natural. If now we lose microscopic reversibility, albeit very weakly, we justified the reverse sequence in terms of recurrence, i.e., because there are only two bodies involved, the recurrence time is relatively short. But did the billiard balls return to the exact original state? As macroscopic objects, the billiard balls look exactly the same and thus we say that the balls went back to their original states (initial positions and velocities). But the discussions in Sections II, III and IV show that we do not get back the exactly same billiard balls because of the changes that happened to some of its atoms and nuclei as a result of the interaction with the environment and the time-reversal violating interactions due to the electric dipole moment of the neutrons and electrons. Thus strictly speaking, what appears as a reverse collision is evolving forward in time for two reasons - recurrence and the balls are not exactly the same as when they had the original collision.
	
	The picture that emerges from these discussions is that (1) the internal state of the elements of the thermodynamioc system changes, although imperceptibly, in an irreversible manner due to CP violation, and (2) the irreversible evolution of the universe is like a flowing river that makes everything in it flow forward and in the process increase entropy. But how do we explain biological systems (during growth stage, it can lower its entropy) that seem to violate the flow, at least for a brief moment compared to the long time processes in the universe. The biological system has a mechanism, in this case photosynthesis, which allows it to harness the high entropy sunlight to produce a more ordered accumulation of atoms and molecules (this is where the H Theorem, original and modified, will fail primarily because the assumption of Boltzmann no longer holds). The analogy is that this system is like a fish in a river that is able to get nutrients from the river and able to swim against the flow for a certain period. But eventually, the universe wins - all in the universe will irreversibly flow towards the final state (whatever it is, Big Freeze or Big Rip), just like the river eventually wins - everything in it ends in the sea. 
\begin{acknowledgments}
The author appreciates the comments of the referee, which made him think about the problem more carefully. The author would like to thank the Creative Work and Research Grant of the University of the Philippines System for supporting his research.
\end{acknowledgments}

\end{document}